\documentclass[12pt]{article}

\usepackage{amsmath}
\usepackage{amssymb}
\usepackage{braket}

\usepackage{graphicx}
\usepackage{subfigure}
\usepackage{float}
\usepackage{wrapfig}

\usepackage[linesnumbered, ruled]{algorithm2e}

\usepackage[table,xcdraw]{xcolor}
\usepackage{multirow}

\usepackage{booktabs}

\usepackage{cite}
\usepackage{authblk}

\usepackage[english]{babel}
\usepackage{comment}
\usepackage{rotating}
\usepackage{hyperref}  

\date{}

\date{\today}

\title{\bf Machine-Learning-Enhanced Entanglement Detection Under Noisy Quantum Measurements}
\author[1,2,3]{Mahmoud Mahdian\thanks{mahdian@tabrizu.ac.ir}}
\author[4]{Ali Babapour-Azar\thanks {babapoor@tabrizu.ac.ir}}
\author[1,3]{Zahra Mousavi\thanks {z.mosavi1400@ms.tabrizu.ac.ir}}

\author[4,5]{Rashed Khanjani-Shiraz\thanks {r.khanjani@tabrizu.ac.ir} }
\date{\today}

\affil[1]{Faculty of Physics, Theoretical and Astrophysics
	Department, University of Tabriz, 51665-163 Tabriz, Iran}
\affil[2]{Research Institute for Applied Physics and Astronomy (RIAPA), University of Tabriz, Tabriz, Iran}
\affil[3]{Quantum Technology Center, University of Tabriz, Tabriz, Iran}
\affil[4]{School of Mathematics, University of Tabriz, Tabriz, Iran}
\affil[5]{Department of Engineering and Economics,South Westphalia University of Applied Sciences, Germany}

\begin{document}
\maketitle
\begin{abstract}
	
Quantum measurements are inherently noisy and data-intensive, posing significant challenges for reliable entanglement detection and the scalability of quantum technologies. While error mitigation techniques exist, they often require a prohibitive number of measurements, making the process resource-intensive. In this work, we demonstrate that by explicitly accounting for measurement errors, high-fidelity entanglement detection can be achieved with significantly less data.
We introduce a machine-learning-based approach that delivers noise-resilient entanglement classification even with imperfect measurements. Our method employs support vector machines (SVMs) trained on features from Pauli measurements to construct a robust optimal entanglement witness (ROEW). By optimizing SVM parameters against worst-case errors, our protocol ensures effectiveness under unknown measurement noise. Numerical experiments show that ROEW maintains high classification accuracy even when measurement errors exceed 10\%. Crucially, we demonstrate that training the model using only 20\% of the typical dataset suffices to achieve high accuracy and substantial error reduction. Our proposed ROEW significantly outperforms traditional non-robust models, maintaining superior detection performance under elevated noise. This work bridges machine learning and quantum information science, offering a practical tool for noise-robust quantum characterization and advancing the feasibility of entanglement-based technologies.

\end{abstract}

\section{Introduction}

Errors in quantum circuits caused by quantum measurements arise from the probabilistic nature of quantum mechanics and practical limitations in measurement processes \cite{shor1996fault}. Quantum measurements collapse the superposition state of qubits, often disrupting entanglement and causing decoherence or unintended state changes \cite{chuang1995quantum,wiseman2009quantum}. These errors are exacerbated by environmental noise, hardware imperfections, inaccuracies in state detection, signal loss, calibration drift, and limited qubit coherence times \cite{terhal2015quantum,preskill1998reliable}. Specific types of errors include measurement noise from misreporting qubit states, state-dependent bias due to decay during measurement, and correlated errors that affect multiple qubits simultaneously \cite{braginsky1995quantum,unruh1995maintaining}. To mitigate these issues, researchers employ error correction codes that use additional qubits to detect and fix errors, measurement error mitigation (MEM) techniques like Matrix-free Measurement Mitigation (M3) \cite{PRXQuantum.2.040326} to recalibrate data, and dynamic error adjustment strategies such as Active Inferential Measurement (AIM) \cite{PhysRevA.105.062404} that adapt to device error profiles. Improving hardware precision and minimizing environmental interference are also critical for maintaining coherence and reducing measurement disturbances, enabling more reliable and scalable quantum computing systems \cite{bennett1996mixed}.\\
Quantum entanglement, a cornerstone of quantum computing and information processing, enables applications such as quantum algorithms, cryptography, teleportation \cite{Nielsen, Bennett,Bennett1,PhysRevLett.67.661}, and dense coding through its non-classical correlations \cite{Einstein,Schrodinger,Nielsen}. Detecting entanglement in high-dimensional quantum states is essential due to the increasing complexity of multi-partition systems \cite{Horodecki}, and various methods for detecting entanglement, such as entanglement witnesses(EWs) \cite{mahdian2025entanglement,mahdian2025optimal,Jafarizadeh,PhysRevA.78.032313}, Bell inequalities, state tomography, and positive partial transpose (PPT) criteria, are employed to identify and quantify it. Advanced techniques like machine learning-based approaches and tensor network methods are also gaining prominence for handling the complexity of high-dimensional systems. These methods enable researchers to efficiently characterize entanglement, which is critical for optimizing quantum protocols and ensuring the scalability of quantum technologies.\\
First, we recall the definition of pure and mixed separable and entangled states, as well as EWs.

\emph{\textbf{Definition 1: Separability in Multipartite Systems}}

A pure state $\ket{\psi}$ of an $n$-partite quantum system is called \textbf{$j$-separable} ($2 \leq j \leq n$) if it can be decomposed into a tensor product of $j$ subsystem states:

\[
\ket{\psi} = \ket{\alpha_1} \otimes \ket{\alpha_2} \otimes \dots \otimes \ket{\alpha_j},
\]

where each $\ket{\alpha_i} \in \mathcal{H}_{d_i}$ is a state in Hilbert space of dimension $d_i$, and the composite system satisfies $\prod_{i=1}^j d_i = d$ (where $d$ is the total dimension of the system). The corresponding density matrix is:

\[
\rho_{\text{sep}}^j = \ket{\alpha_1}\bra{\alpha_1} \otimes \ket{\alpha_2}\bra{\alpha_2} \otimes \dots \otimes \ket{\alpha_j}\bra{\alpha_j}.
\]

For \textbf{mixed states}, $\rho$ is fully separable if it can be written as:

\[
\rho = \sum_i p_i \rho_i^{(1)} \otimes \rho_i^{(2)} \otimes \dots \otimes \rho_i^{(n)}, \quad \text{with} \quad \sum_i p_i = 1, \ p_i \geq 0.
\]

If no such decomposition exists, $\rho$ is entangled \cite{Horodeckii,Gabriel}.

\emph{\textbf{Definition 2: Entanglement Witness}}

EWs are Hermitian operators that distinguish entangled states from separable ones by acting as hyperplanes, leveraging the convexity of separable states as guaranteed by the Hahn-Banach theorem \cite{guhne2009entanglement}. An EW on $\mathcal{H}$ satisfying:

\begin{enumerate}
	\item $\text{Tr}(W\rho_{\text{sep}}) \geq 0$ for all separable $\rho_{\text{sep}}$.
	\item There exists at least one negative eigenvalue and $\text{Tr}(W\rho_{\text{ent}}) < 0$ .
\end{enumerate}
The use of machine learning methods and different algorithms in solving quantum information and computing has grown a lot in recent years. The SVM algorithm is used for classification, where different classes can be separated from each other with a hyperplane. This method can be used to distinguish the class of separable states from entangled states. Using method SVM, we can obtain EWs that are completely on the boundary of the separable tangent area and are optimal witnesses \cite{svmlearning,support}.
Entanglement detection methods based on machine learning require measurements, and as we know, measurements in physics are always accompanied by a degree of uncertainty or error that must be explicitly accounted for in the measured quantities. \\
Robust optimization addresses data uncertainty by constraining variations within a predefined deterministic set, aiming to find optimal solutions under worst-case scenarios. In contrast, Distributionally Robust Optimization (DRO) considers probabilistic uncertainty, seeking to minimize expected loss over a set of possible distributions derived from observed data. This approach is particularly beneficial when training data is limited or noisy, as traditional classifiers may struggle to generalize effectively. By employing a generalized DRO framework, we aim to enhance the performance of SVMs under such conditions.
Extensive research has been conducted on Distributionally Robust Chance-Constrained (DRC) linear SVM models, with significant contributions from scholars like Ben-Tal et al. \cite{ben2011chance}, Wang et al. \cite{wang2018robust}, Khanjani-Shiraz et al. \cite{khanjani2023distributionally}, and Faccini et al. \cite{faccini2022robust}. These studies highlight the depth of research in this area, demonstrating its importance in addressing uncertainty in classification problems.
This work introduces a machine-learning-enhanced framework for entanglement detection that remains robust under significant measurement noise. Our approach leverages SVMs trained on features extracted from Pauli measurements. To address inherent experimental uncertainties in the training data—arising from limitations in the measurement process and instrument precision—we employ a robust-SVM method. This technique directly incorporates data uncertainties into the analysis, mitigating the impact of noise and errors during training.
The core outcome of this method is a ROEW, which reliably classifies quantum states even in the presence of unknown and substantial measurement imperfections. This overcomes a key limitation of traditional witness operators, which typically assume ideal or well-characterized measurement settings.
A significant advantage of our framework is its high data efficiency. Numerical simulations demonstrate that training the model with only 20\% of the available dataset is sufficient to achieve high classification accuracy and substantial error reduction. This efficiency is favorable compared to many existing ML-based or conventional protocols that require larger, more calibrated datasets.
Furthermore, our ROEW substantially outperforms conventional non-robust models, maintaining superior detection performance across a wide range of noise levels—including cases where measurement errors exceed 10\%. This resilience is crucial for real-world applications where noise characteristics are unknown or dynamically varying. The model also retains significant accuracy even at lower confidence thresholds, enabling flexible deployment where trade-offs between confidence and resource consumption are necessary.
Collectively, these features establish our approach as a practical and salable tool for entanglement detection in noisy quantum environments. By bridging advances in machine learning and quantum information science, this work paves the way for enhanced quantum state characterization and supports the development of noise-resilient quantum technologies.

This paper is organized as follows:  Section 2 introduces the foundational SVM models and reviews relevant literature. In Section 3, we examine the principles of distributional robustness within the SVM framework. Section 4 details the core ideas, mathematical programming models, and supporting theorems. Section 5 presents and analyzes the numerical results. Finally, Section 6 concludes the paper with a summary of findings and potential directions for future research.

\section{Preliminaries and general definitions}

SVM, as introduced by Vapnik in 1979 and further developed by Boser et al. in 1992, stand out as widely embraced and effective classification methods. SVM hinge on a fundamental concept: discovering a hyperplane within the feature space that maximizes the margin between data points. In cases where samples are not linearly separable, Cortes and Vapnik (1995) introduced the soft-margin SVM. This adaptation includes a penalty term for each non-separable sample in the objective function. SVMs are now widely accepted as effective methods for classification. The key idea is to identify a hyperplane that maintains the greatest possible separation between classes. In non-linearly separable scenarios, the soft-margin SVM addresses the challenge by adding a penalty term for misclassified samples, leading to a better functional performance in practical applications.
Recently, Khanjani et al. (2023) proposed a distributionally robust chance-constrained SVM framework, enhancing its structure to account for dependencies among data points and applying copula theory for solving joint chance-constrained SVM issues.

\begin{figure} [H]
	\centering
	\includegraphics[width=0.6\linewidth]{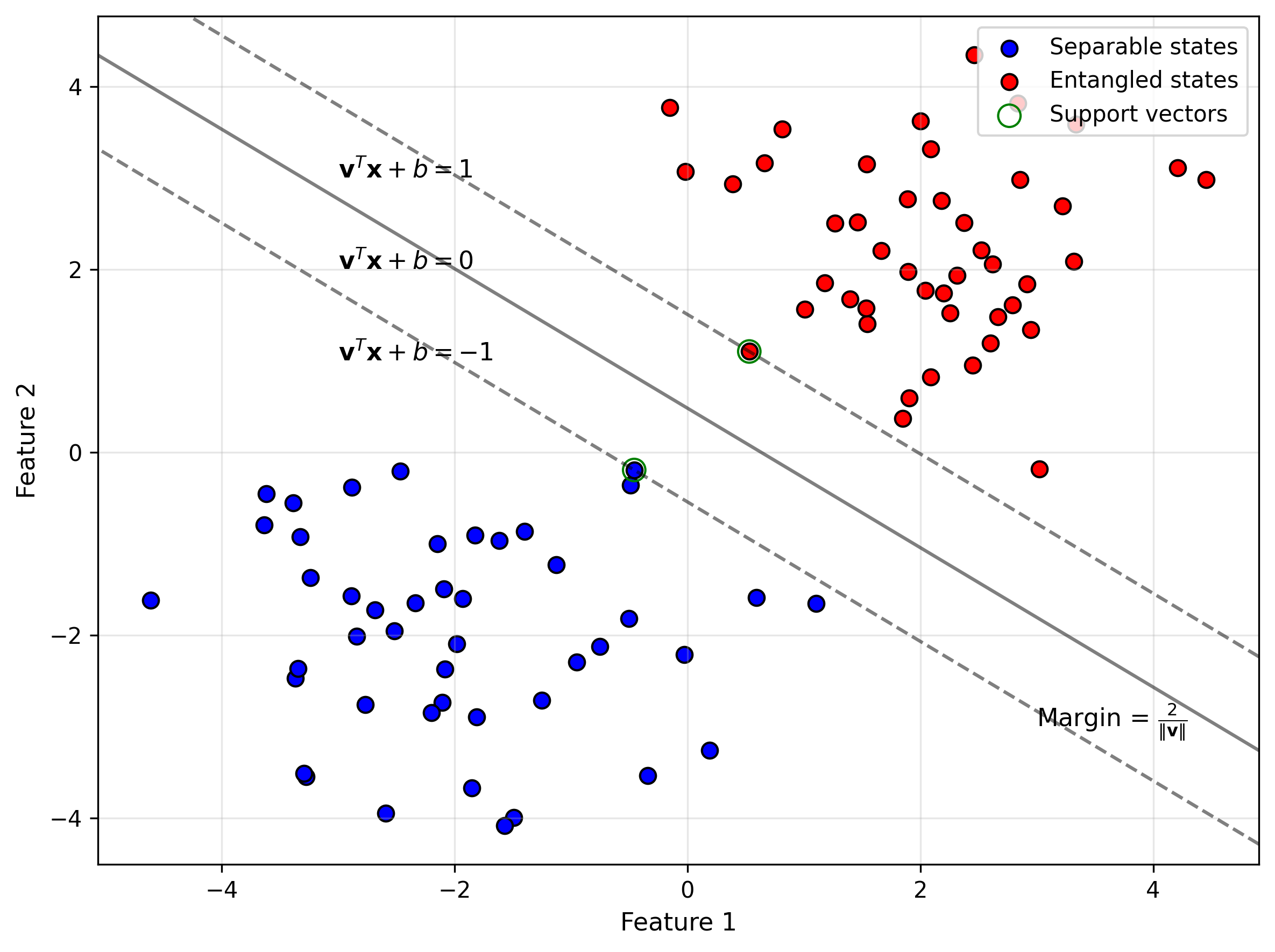}
	\caption{Visualization of a two-dimensional dataset with two classes separated by a hyperplane. Blue circles represent separable states, while red circles denote entangled states. The decision boundary $\mathbf{v}^T\mathbf{x} + b = 0$ (solid line) and margins $\mathbf{v}^T\mathbf{x} + b = \pm 1$ (dashed lines) margin width $2/\|\mathbf{v}\|$ . Support vectors are highlighted in green.}
\label{fig:svm-example}
\end{figure}

In the context of binary classification, training data is denoted by 
	. Our goal is to determine a linear classifier represented by $f(x; v, b) = \text{sign}(v^Tx + b)$. The SVM algorithm aims to define a hyperplane characterized by $( \mathbf{v}, b)$ that maximizes the margin, quantified as $\frac{1}{\|\mathbf{v}\|}$, where $\|\cdot\|$ indicates the Euclidean norm.
Figure \ref {fig:svm-example} illustrates a two-dimensional dataset comprising two classes, denoted by circles and triangles, separated by a hyperplane.

For data that is not linearly separable, the soft-margin SVM attempts to find the vector $\mathbf{w}$ and scalar $b$ through the following quadratic optimization problem:
\begin{equation}
	\begin{aligned}
		\text{minimize} \quad & \frac{1}{2}\mathbf{v}^T \mathbf{v} + C \sum_{j=1}^{m} s_j ,\\
		\text{subject to} \quad & s_j \geq 1 - y_j(\mathbf{v}^T \mathbf{x}_j + b), \quad (j = 1, \ldots, m), \\
		& s \geq 0.
	\end{aligned}
\end{equation}

where $C$ is a regularization parameter, and $s_j$ are slack variables that allow for some degree of misclassification while enforcing classification accuracy.

\subsection{Distributionally Robust Optimization (DRO)}
Robust optimization is a method for managing uncertainty in optimization without relying on probabilistic assumptions. It constrains data variations to a predefined deterministic set, aiming for solutions that perform satisfactorily under worst-case scenarios. In contrast, DRO handles uncertainty with a probabilistic lens, minimizing expected loss in the worst-case scenario over a set of possible distributions that have known properties (Ben-Tal et al., 2011; Wang et al., 2018).

DRO primarily utilizes two types of ambiguity sets: moment-based sets, which include distributions adhering to specific moment constraints, and statistical distance sets, which account for variations from empirical distributions.

Moment-based ambiguity sets, defined by specific moment constraints, offer enhanced tractability. In this paper, we proceed under the assumption that the first and second-order moments of the random variable \(x_i\) are given. We define the mean and covariance matrix of this random variable as follows:

\begin{equation}
	\begin{aligned}
		\mu_i &= \mathbb{E}[x_i], \quad  \\
		\Sigma_i &= \mathbb{E}[(x_i - \mu_i)(x_i - \mu_i)^T], \quad (i = 1, \ldots, K),
	\end{aligned}
\end{equation}
where \(x_i = [x_{i1}, \ldots, x_{ik}]^T\).
\subsection{Distributionally Robust Support Vector(DRSV) }

In quantum computing, measurements are essential yet prone to errors, impacting the computations. The outcome of a measurement is not always the same even under identical conditions, leading to noise in the measurement.Addressing measurement error is crucial for the progress of quantum computing.  
We address this by employing DRO and SVM to refine the accuracy of EW measurements, thereby substantially improving the reliability of outcomes in quantum computing tasks.

With strong assumptions about the mean and covariance, we can solve the robust SVM problem.

DRO helps us deal with uncertainties and unexpected changes during the measurement process, making our method robust and reliable. On the other hand, SVM helps us organize and analyze the complex data from quantum computing in a very structured way.
By combining DRO and SVM, we greatly improve how well we can measure quantum entanglement with witness operators. This improvement is a big step in making quantum computing results more reliable and trustworthy.
The Chance-Constrained Program (CCP) focuses on minimizing classification errors while considering data uncertainties. The chance-constrained SVM is defined as:

\begin{equation}
\begin{aligned}
\min \quad & \frac{1}{2} v^T v + C \sum_{i=1}^{K} \beta_i, \\
\text{s.t.} \quad & \ \text{prob}[\lambda_i(v^T x_i + b) \geq 1 - \beta_i] \geq \alpha, \quad  (i = 1, \ldots, K), \\
& \beta_i \geq 0, \quad  (i = 1, \ldots, K).
\end{aligned}
\end{equation}

Here, $0 < \alpha < 1$ is a parameter defining the confidence level, and $\text{prob}\{\cdot\}$ indicates the probability distribution. When the specific probability distributions of random variables are unknown, the distributionally robust chance constraint presents a conservative approximation to manage uncertainties effectively.
Consequently, the distributionally robust chance-constrained SVM formulation, as proposed by Shivaswamy et al. (2006) and \cite{ben2011chance}, is defined as:

\begin{equation}
\begin{aligned}
\min \quad & \frac{1}{2} v^T v + C \sum_{i=1}^{K} \beta_i, \\
\text{s.t.} \quad & \inf_{x_i \sim (\mu_i, \Sigma_i)} \text{prob}[\lambda_i(v^T x_i + b) \geq 1 - \beta_i] \geq \alpha, \quad (i = 1, \ldots, K), \\
& \beta_i \geq 0, \quad (i = 1, \ldots, K).
\end{aligned}
\end{equation}

The study of distributionally robust chance-constrained SVM models has received substantial attention, with contributions from Ben-Tal et al. (2011), Wang et al. (2018), Khanjani et al. (2023), and Faccini et al. (2022), all of which emphasize the importance of addressing uncertainty in classification tasks while maintaining high accuracy in practical applications.

Formulation of the joint chance-constrained SVMs through semidefinite and second-order cone programming, as outlined below.

\begin{align}
\min \quad & \frac{1}{2} v^T v + C \sum_{i=1}^{K} \beta_i, \nonumber \\
\text{s.t.} \quad & \delta_i - \frac{1}{1 - \alpha} \text{tr}(\phi_i M_i) \geq 0, \quad (i = 1, \ldots, K), \nonumber \\
& M_i + 
\begin{bmatrix}
	\frac{1}{2 \lambda_i} v \\
	\frac{1}{2 \lambda_i} v^T \lambda_i b + \beta_i - 1 - \delta_i
\end{bmatrix} 
\succeq 0, \quad (i = 1, \ldots, K), \nonumber \\
& M_i \succeq 0, \quad (i = 1, \ldots, K), \nonumber \\
& \beta_i \geq 0, \quad  (i = 1, \ldots, K), \ \\
\text{where} \quad & \phi_i = 
\begin{bmatrix}
	\psi_i + \mu_i \mu_i^T & \mu_i \\
	\mu_i^T & 1
\end{bmatrix},
\quad (i = 1, \ldots, K 
).\nonumber
\end{align}

The SOCP model is as follows:
\begin{align}
\min \quad & \frac{1}{2} v^T v + C \sum_{i=1}^{K} \beta_i, \nonumber \\
\text{s.t.} \quad & -\lambda_i (v^T \mu_i + b) \leq \beta_i - 1 - \sqrt{\frac{\alpha}{1 - \alpha}} \left\| \sqrt{\psi_i} v \right\|, \quad (i = 1, \ldots, K), \nonumber \\
& \beta_i \geq 0, \quad  (i = 1, \ldots, K).
\end{align}

\section{Application of DRO in entanglement detection}
In recent years, machine learning has been increasingly applied to solve quantum computing problems, including identifying entangled states. Supervised and unsupervised algorithms like neural networks and SVMs have been used for this purpose. Given that separable states form a convex set describable by a hyperplane, this work aims to separate entangled from separable states using the SVM method.
We consider an EW, a Hermitian operator with at least one negative eigenvalue, used to detect entanglement in a multipartite Hilbert space as follows:

\begin{equation}\label{eq:state}
	W=\sum _{k_1,k_2,...k_n} \chi_{k_1,k_2...k_n} \\ \mathcal{O}_{k_1}\otimes\mathcal{O}_{k_2}...\otimes\mathcal{O}_{k_n}.
\end{equation}

We assume that the Hermitian operators $\{\mathcal{O}_{k_i}\}$ act locally on the $k_i$-dimensional subsystems. Here for simplicity, we will work on the multi-qubit systems, so $\{\mathcal{O}_{k_i}\}$ operators are replaced by Pauli operators $(\sigma_x,\sigma_y,\sigma_z)$ \textcolor{yellow}. First, to use SVM method, we must generate the training data, which are the separable and entangled states. Then we measure the value of Hermitian operators or the expectation values for each of these quantum states which these measurement quantities are always accompanied by a certain amount of uncertainty.

\begin{equation}\label{eq:state1}
	\langle\mathcal{O}_{k_1}\otimes\mathcal{O}_{k_2}...\otimes\mathcal{O}_{k_n}\rangle = tr(\rho \\ \mathcal{O}_{k_1}\otimes\mathcal{O}_{k_2}...\otimes\mathcal{O}_{k_n})
	,
\end{equation}
and
\begin{equation}\label{eq:states}
	\langle\mathcal{O}_{k_i} \rangle = tr(\rho \\ \mathcal{O}_{k_i})=	
	x_i+\delta x_i ,
\end{equation}
where $\delta x_i$ is related to post-measurement noise and $	x_i+\delta x_i $ is the ‘features’ used to train the SVM. It should be noted that the expectation values are always accompanied by errors $\vec{X}+\delta X=(x_1+\delta x_1,x_2+\delta x_2,...,x_n+\delta x_n)$. In this work we include the errors in the values of features using the Robust SVM to obtain the ROEW.

Then we can write:

\begin{equation}
	\langle W\rangle= tr(\rho W),	
\end{equation}

\begin{equation}\label{eq:state2}
	\langle W\rangle= \sum \chi_{k_1,k_2...k_n} (\vec{X}+\delta X)= \sum \chi_{k_1,k_2...k_n} (x_1+\delta x_1,x_2+\delta x_2,...,x_n+\delta x_n),
\end{equation}

\[  
\text{If the $\rho$ is separable,} \quad \langle W\rangle \geq 0  
,\]  

\[  
\text{If the $\rho$ is entangled,} \quad \langle W\rangle < 0  
.\]  

The coefficients $\chi_{k_1,k_2...k_n}$ are learned using training data and labels, label is $y=+1$ for separable states and $y=-1$ for entangled states. By minimizing the following loss function, the best hyperplane or EW can be obtained to distinguish entangled states from separable class. The training data for the entangled class includes features $\vec{X}(\rho _{ent})= tr(\rho _{ent}.\mathcal{O}_{k_1}\otimes\mathcal{O}_{k_2}...\otimes\mathcal{O}_{k_n})$ and label $y=-1$. Also, for the separable class consists of the label $y=+1$ and features $\vec{X}(\rho _{sep})= tr(\rho _{sep}.\mathcal{O}_{k_1}\otimes\mathcal{O}_{k_2}...\otimes\mathcal{O}_{k_n})$.
We define the loss function $\mathcal{L}$, which combines a margin-based classification error with $L_1$ regularization, as follows:
\begin{equation}\label{eq:state3}
\mathcal{L}=\frac{1}{N}  \sum_{i=1}^N \{max(0,1- \hat{\textbf{y}}^i.y^i)\}^m+C \sum | \chi_{\vec{k}}| ,
\end{equation}
The first term computes the $m$-th power of the hinge loss across $N$ samples, where $\hat{\mathbf{y}}^i$ represents the predicted output and $y^i$ the true label for sample $i$. The second term adds $L_1$ regularization on parameters $\chi_{\vec{k}}$ with strength $C$, promoting sparsity in the solution. This formulation enables robust classification while performing feature selection.

Bell-Diagonal states are two-qubit states diagonal in the Bell basis:
\begin{equation}
	\rho_{\text{BD}} = \sum_{i=1}^4 p_i |\Psi_i\rangle\langle\Psi_i|,
\end{equation}
where $\{|\Psi_i\rangle\} = \{|\Phi^+\rangle, |\Phi^-\rangle, |\Psi^+\rangle, |\Psi^-\rangle\}$ are the Bell states and $\sum_i p_i = 1$.

\begin{figure}[H]
	\centering
	\includegraphics[scale=0.33]{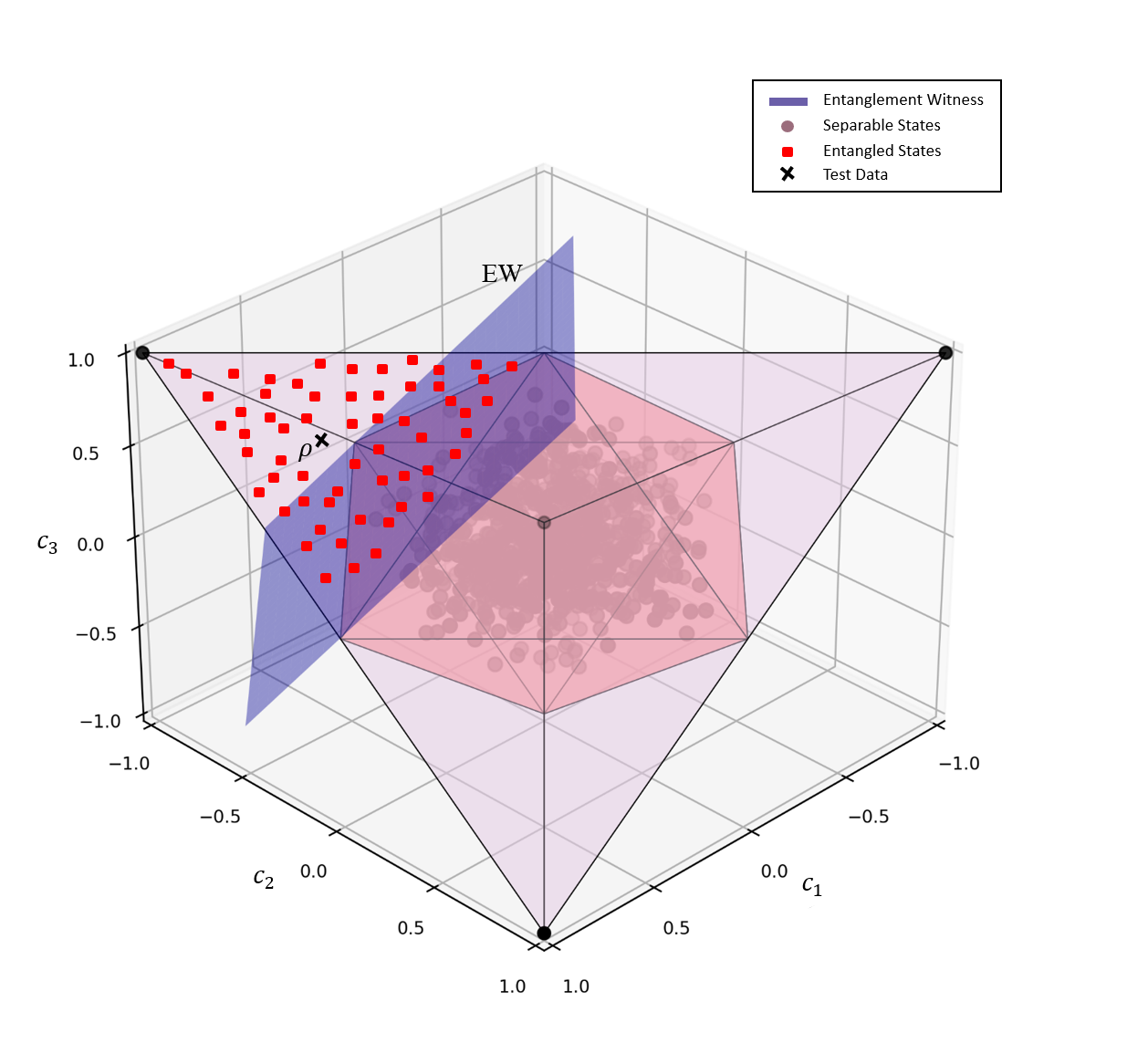}
	\caption{This 3D plot illustrates a robust chance-constrained SVM classifier applied to two-qubit
		Werner density matrices for distinguishing entangled from separable quantum states. The
		axes $c_1$, $c_2$ and $c_3$ represent features derived from the density matrix. The purple region
		contains separable states (shown as light purple dots), while the red squares represent
		entangled states. A dark blue plane labelled "EW" depicts the entanglement witness, which
		acts as the decision boundary learned by the robust SVM classifier. The black cross marked
		$\rho^{\textbf{x}}$ is a test quantum state, whose position relative to the separating plane indicates its classification. This visualization demonstrates the ability of the robust SVM to accurately separate quantum states under uncertainty.}
	\label{fig:c11}
\end{figure}

In our analysis to construct data of entangled and separable states, we employ Werner states – a special one-parameter subclass of Bell-diagonal states that remain invariant under arbitrary $U\otimes U$ unitary transformations::
\begin{equation}
	\rho_W = \gamma|\Psi^-\rangle\langle\Psi^-| + \frac{1-\gamma}{4}\mathbb{I}_4, \quad \gamma \in [-1/3,1],
\end{equation}
where entanglement occurs when $\gamma > 1/3$. These are special cases of Bell-diagonal states with $p_1 = p_2 = p_3 = (1-\gamma)/4$ and $p_4 = (1+3\gamma)/4$.(Fig.\ref{fig:c11})

\subsection{Designing an SVM-Based Algorithm}
In this section we outline the objectives and structure of our proposed SVM algorithm for  ROEW, highlighting its innovations and contributions to existing methodologies.
Our aim is to present an algorithm that provides hyperplanes such that the separable region lies within the intersection of these hyperplanes.

\begin{algorithm}[H]
	\caption{Robust Optimal Entanglement Witness}\label{alg:ROEW}
	\textbf{Requirement:} $\alpha$ is a confidence level parameter. \\
	\textbf{Input:} A penalty parameter $C$ and training data set $\{x,y\}$. \\
	\textbf{Step 1:} Divide the training dataset belonging to the entangled group into 4 groups, each representing one of the Bell states. Then, name all samples belonging to each of the states $|\Phi^+\rangle$, $|\Phi^-\rangle$, $|\Psi^+\rangle$, and $|\Psi^-\rangle$ respectively as $x^{00}$, $x^{01}$, $x^{10}$, and $x^{11}$. \\
	\textbf{Step 2:} 
	\For{$\mu$ in $I=\{00, 01, 10, 11\}$}{
		Solve following optimization problem:
		\begin{align}\label{eq:en}
			\min \quad & \frac{1}{2} v^T v + C \sum_{i=1}^{K} \beta_i, \nonumber \\ 
			\text{s.t.} \quad & -(v^T \mu_j + b) \leq -1 - \sqrt{\frac{\alpha}{1 - \alpha}} \left\| \sqrt{\psi_j} v \right\|, \nonumber \\ 
			& \quad (j = 1, \ldots, K_{sep}), \nonumber \\ 
			& (v^T \mu_j^{i} + b) \leq \beta_j - 1 - \sqrt{\frac{\alpha}{1 - \alpha}} \left\| \sqrt{\psi_j^{i}} v \right\|, \nonumber \\  
			& \quad (j = 1, \ldots, K_{en}), \nonumber \\
			& \beta_i \geq 0, \quad (i = 1, \ldots, K_{en}).
		\end{align}
		Let $v_{\mu}^{opt}$ and $b_{\mu}^{opt}$ be the optimal solutions of \eqref{eq:en}. \\
		Set $\chi_{\mu} = (b_{\mu}^{opt}, v_{\mu}^{opt})$.
	}
	\textbf{Step 3:} Construct witness operator:
	\begin{equation}
		EW_{\eta} = \sum_{i_1,...,i_k} \chi_{i_1,...,i_k} \sigma_{i_1}  \otimes \sigma_{i_k} , \quad \eta=[1,...,4]  , \quad i_k=\{I,\sigma_x,\sigma_y,\sigma_z\}.
	\end{equation}
	
	Set $EW = [EW_1, EW_2, EW_3, EW_4]$ for Bell states. \\
	\textbf{Output:} $EW$ 
	
\end{algorithm}

\section{Computational Study}

To evaluate the effectiveness of the proposed method, we conducted a comprehensive set of experiments by varying both the proportion of training data and the confidence level parameter~$\alpha$. Specifically, model performance was assessed across four training/testing splits: 20\%, 40\%, 60\%, and 80\% of the data used for training, with the remaining portion reserved for testing.

For each data split, we systematically varied $\alpha$ in the range of 0.55 to 0.95. This parameter represents the confidence level, which controls the model's decision threshold—higher values of $\alpha$ enforce stricter criteria for classification, potentially improving precision at the cost of recall.

The performance of the proposed model was evaluated using three widely adopted classification metrics: accuracy, precision, and F1-score. The results, corresponding to different confidence levels ($\alpha$) and varying proportions of training data, are comprehensively summarized in Table~\ref{tab:1}.

Overall, the experimental results confirm the effectiveness and robustness of the proposed method across a wide range of confidence levels and training data ratios. Notably, the method maintains high accuracy, precision, and F1-score even when trained on a small fraction of the dataset, demonstrating its ability to generalize well with limited supervision.

Such performance suggests that the approach is particularly well-suited for real-world applications where access to large-scale annotated datasets is restricted. By reducing the dependency on extensive training data, the method offers a practical and scalable solution for data-scarce environments.
\begin{table}[H]
    \centering

    \scalebox{0.9}{
    \begin{tabular}{c c c c c}
        \toprule
        \textbf{$\alpha$} & \textbf{20\% Train} & \textbf{40\% Train} & \textbf{60\% Train} & \textbf{80\% Train} \\
        \midrule
        \multicolumn{5}{c}{\textbf{Accuracy}} \\
        \midrule
        0.55 & 0.99025 & 0.99975 & 1.00000 & 1.00000 \\
        0.60 & 0.99150 & 0.99925 & 0.99975 & 1.00000 \\
        0.65 & 0.99075 & 1.00000 & 1.00000 & 1.00000 \\
        0.70 & 0.99050 & 1.00000 & 1.00000 & 1.00000 \\
        0.75 & 0.99325 & 1.00000 & 1.00000 & 1.00000 \\
        0.80 & 0.99650 & 1.00000 & 1.00000 & 1.00000 \\
        0.85 & 0.99650 & 0.99975 & 1.00000 & 1.00000 \\
        0.95 & 0.99875 & 1.00000 & 1.00000 & 1.00000 \\
        \midrule
        \multicolumn{5}{c}{\textbf{Precision}} \\
        \midrule
        0.55 & 0.98050 & 0.99950 & 1.00000 & 1.00000 \\
        0.60 & 0.98300 & 0.99850 & 0.99950 & 1.00000 \\
        0.65 & 0.98150 & 1.00000 & 1.00000 & 1.00000 \\
        0.70 & 0.98100 & 1.00000 & 1.00000 & 1.00000 \\
        0.75 & 0.98650 & 1.00000 & 1.00000 & 1.00000 \\
        0.80 & 0.99300 & 1.00000 & 1.00000 & 1.00000 \\
        0.85 & 0.99300 & 0.99950 & 1.00000 & 1.00000 \\
        0.95 & 0.99750 & 1.00000 & 1.00000 & 1.00000 \\
        \midrule
        \multicolumn{5}{c}{\textbf{F1-score}} \\
        \midrule
        0.55 & 0.99015 & 0.99975 & 1.00000 & 1.00000 \\
        0.60 & 0.99143 & 0.99925 & 0.99975 & 1.00000 \\
        0.65 & 0.99066 & 1.00000 & 1.00000 & 1.00000 \\
        0.70 & 0.99041 & 1.00000 & 1.00000 & 1.00000 \\
        0.75 & 0.99320 & 1.00000 & 1.00000 & 1.00000 \\
        0.80 & 0.99649 & 1.00000 & 1.00000 & 1.00000 \\
        0.85 & 0.99649 & 0.99975 & 1.00000 & 1.00000 \\
        0.95 & 0.99875 & 1.00000 & 1.00000 & 1.00000 \\
        \bottomrule
    \end{tabular}
    }
     \caption{Model performance metrics (Accuracy, Precision, and F1-score) for varying confidence levels ($\alpha$) and training/testing splits.}
     \label{tab:1}
\end{table}
\begin{figure}[H]
\centering
\includegraphics[scale=0.65]{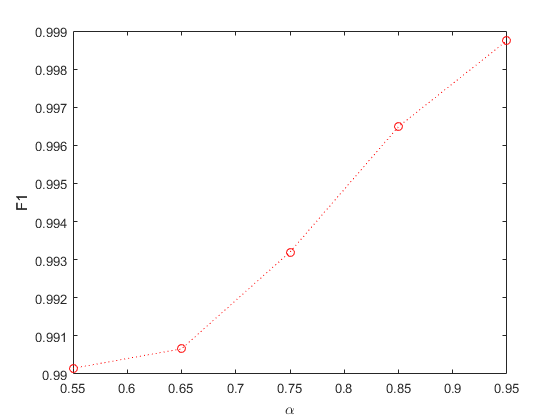}
\caption{Distribution of F1 scores showing model performance across quantum state classifications in  different noisy conditions rate .}\label{c:2}
\end{figure}
Figure \ref{c:2} illustrates the F1 score values across different confidence levels when 20\% of the data was used for training. As shown in the figure, the F1 score improves progressively with increasing confidence levels.

In conclusion, the results  highlight that the model achieves near-perfect performance in accuracy, precision, and F1 score across a variety of training-test splits. These findings suggest that the model is highly reliable and suitable for tasks requiring accurate and balanced performance across multiple evaluation metrics.

To evaluate the effectiveness of the proposed method for entanglement classification under noisy measurement conditions and to investigate its capability to minimize the required measurement data, we generated a dataset consisting of 800 samples. This dataset was partitioned into varying fractions of training samples (20\%, 40\%, 60\%, and 80\%) with the remainder used for testing. For each fraction, the model was trained and subsequently evaluated on an independently generated test set of 4000 samples to rigorously assess its generalization performance and classification accuracy.\\
The results demonstrate that the model achieves high classification accuracy even with a limited fraction of training data, underscoring its ability to extract relevant entanglement features despite the presence of noise. Importantly, this confirms that the proposed approach can significantly reduce the measurement overhead required for reliable entanglement detection. Additionally, the Robust Optimal Entanglement Witness (ROEW) developed in this framework exhibits strong resilience against unknown measurement noise, maintaining robust performance across varying noise levels and training data proportions.

\begin{figure}[H]
    \centering
    \subfigure[0.2]{
        \includegraphics[width=0.45\textwidth]{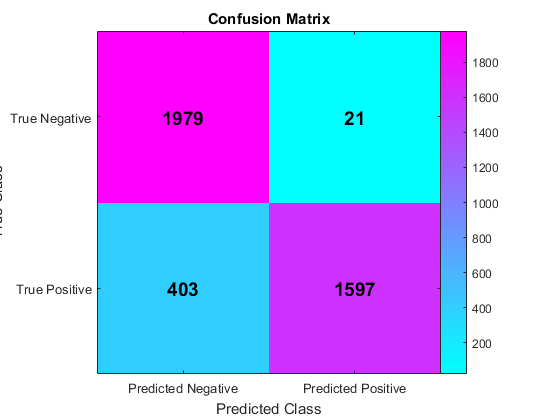}
        \label{fig:subfig1}
    }
    \subfigure[0.4]{
        \includegraphics[width=0.45\textwidth]{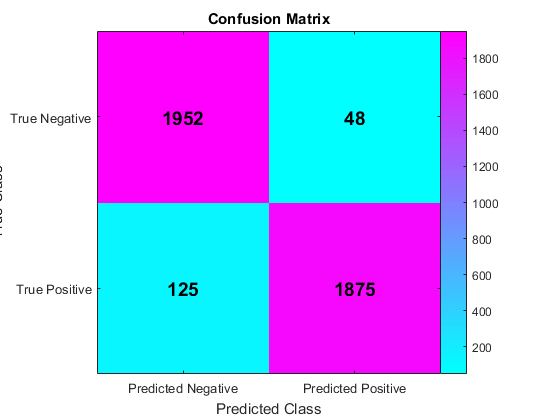}
        \label{fig:subfig2}
    }
    \\
    \subfigure[0.6]{
        \includegraphics[width=0.45\textwidth]{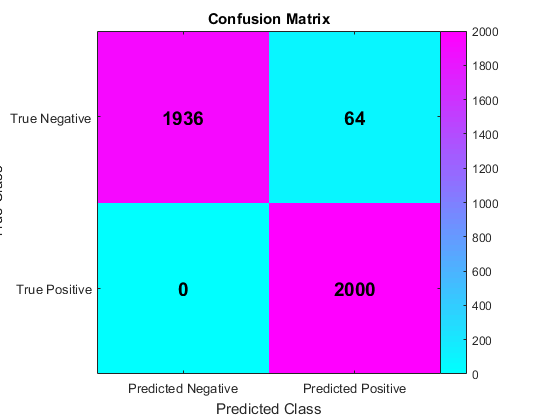}
        \label{fig:subfig3}
    }
    \subfigure[0.8]{
        \includegraphics[width=0.45\textwidth]{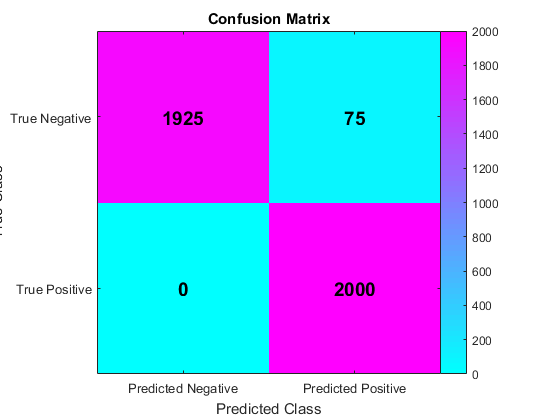}
        \label{fig:subfig4}
    }
    \caption{Confusion matrix for $\alpha = 0.9$.}
    \label{fig:confusion_matrices}
\end{figure}

To gain deeper insights into the model's predictive behavior, confusion matrices for each data percentage are provided in Figure~\ref{fig:confusion_matrices}. These matrices reveal a balanced distribution of true positives and true negatives, demonstrating the model's capability to accurately identify both entangled and separable states. The low number of false positives and false negatives across all cases indicates a strong classification precision and robustness to measurement noise, which is critical in practical quantum information processing tasks.

\begin{figure}[H]
\centering
\includegraphics[scale=0.58]{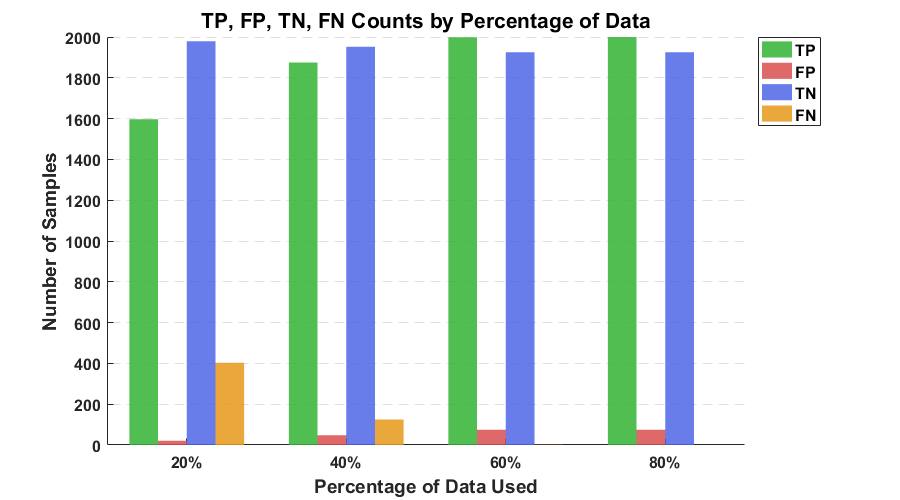}
\caption{TP,TN, FP, FN counts by percentage of data.}\label{fig:tp_fp_tn_fn}
\end{figure}

To complement the statistical metrics, Figure~\ref{fig:tp_fp_tn_fn} illustrates the absolute counts of true positives, false positives, true negatives, and false negatives for each data subset. This granular view highlights the model’s reliability and stability across data volumes, confirming its practical viability for entanglement detection tasks where minimizing misclassification is paramount.

To assess the discriminative ability of the classification model, the Receiver Operating Characteristic (ROC) curve was constructed using the predicted probabilities alongside the actual class labels. The ROC curve depicts the relationship between the true positive rate (sensitivity) and the false positive rate (1 - specificity) across different classification thresholds. The model consistently demonstrated a high true positive rate while maintaining a low false positive rate.

\begin{figure}[H]
    \centering 
    \subfigure[ROC curves for different training data proportions (20\%, 40\%, 60\%, and 80\%) at a fixed confidence level of 0.9.]{
        \includegraphics[width=0.75\textwidth]{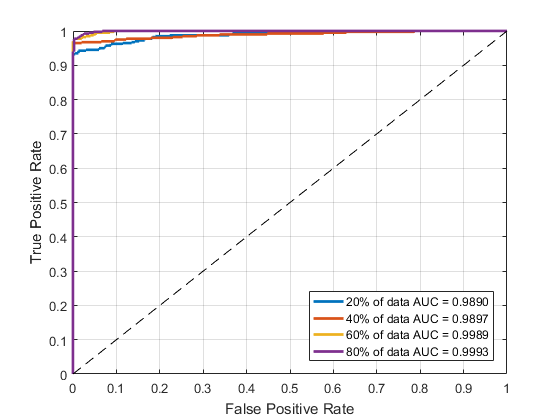}
        \label{fig:2}
    }
    \hfill
    \subfigure[ROC curves for varying confidence levels (0.6, 0.7, 0.8, and 0.9) using 60\% of the training data.]{
        \includegraphics[width=0.75\textwidth]{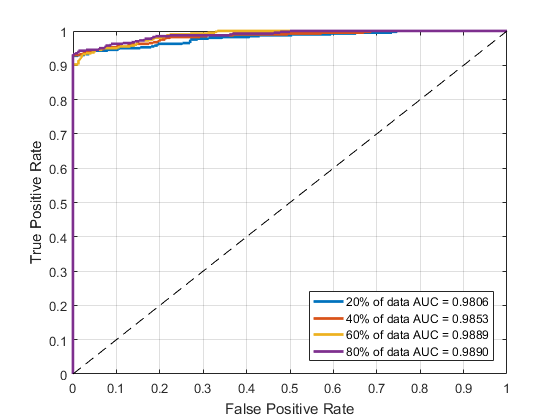}
        \label{fig:3}
    }
    \caption{Performance evaluation under varying conditions.}
    \label{fig:roc_overall}
\end{figure}

Figure~\ref{fig:roc_overall} illustrates the Receiver Operating Characteristic (ROC) curves demonstrating the classification performance of the proposed entanglement detection method under varying conditions. Figure~\ref{fig:2} shows the impact of different training data proportions (20\%, 40\%, 60\%, and 80\%) on the model's accuracy at a fixed confidence level of 0.9. As the amount of training data increases, the model exhibits improved discrimination ability, reflected in higher true positive rates and larger areas under the curve (AUC). The AUC values for the training proportions of 20\%, 40\%, 60\%, and 80\% are 0.9890, 0.9897, 0.9989, and 0.9993, respectively. Figure~\ref{fig:3} explores the effect of varying confidence thresholds (0.6, 0.7, 0.8, and 0.9) on the ROC curves using 60\% training data. Increasing the confidence level generally tightens the classification criteria, influencing the trade-off between sensitivity and specificity. The corresponding AUC values for these confidence levels are 0.9806, 0.9853, 0.9889, and 0.9890, respectively. These results highlight the robustness and flexibility of the proposed method in adapting to different training sizes and confidence requirements.


In addition to the ROC analysis, we further evaluated the performance and reliability of the proposed entanglement classification model through three complementary visualizations.In addition to the ROC analysis, we further evaluated the performance and reliability of the proposed entanglement classification model through three complementary visualizations, all generated using 20\% of the available training data and a fixed confidence level of 0.9.

\begin{figure}[H]
\centering
\includegraphics[scale=0.6]{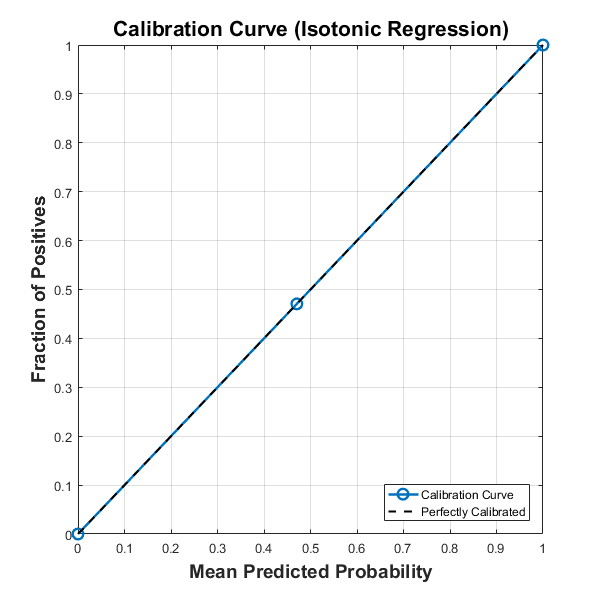}
\caption{calibration curve.}\label{fig:calibration_curve}
\end{figure}

\begin{figure}[H]
\centering
\includegraphics[scale=0.6]{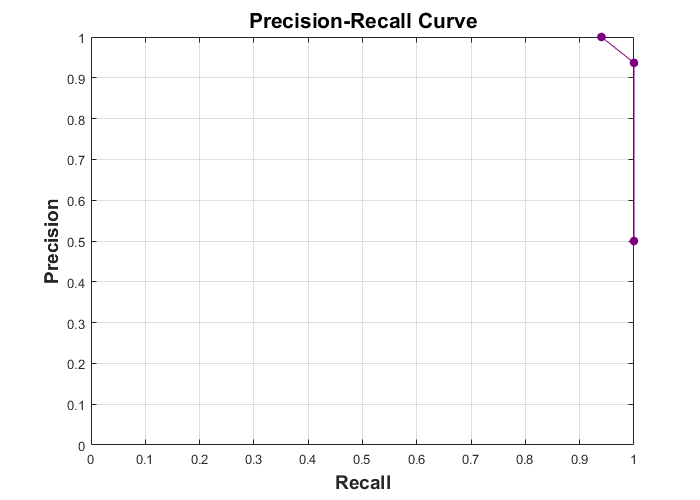}
\caption{Precision, Recall Score for $\alpha = 0.9$.}\label{fig:precision_recall_curve}
\end{figure}

\begin{figure}[H]
\centering
\includegraphics[scale=0.6]{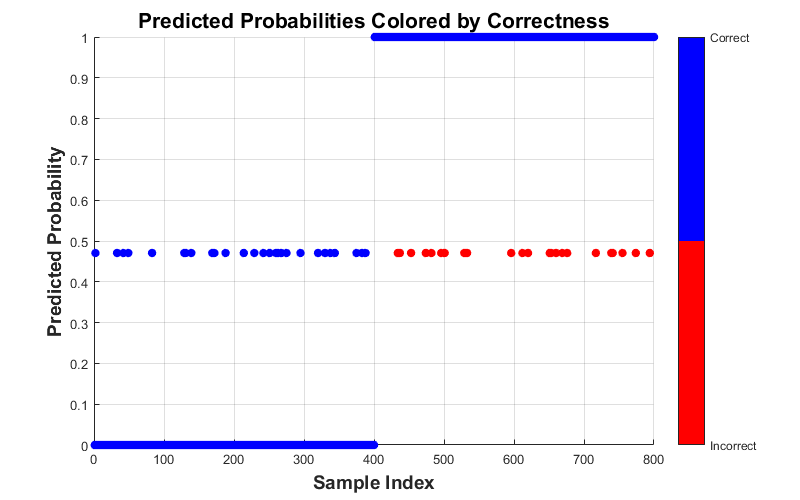}
\caption{ predicted probabilities colored by correctness.}\label{fig:scatter_probabilities}
\end{figure}

First, the calibration curve (Figure~\ref{fig:calibration_curve}) assesses the alignment between the predicted probabilities and the true event frequencies. The observed alignment indicates that the model produces well-calibrated confidence estimates, which is essential for reliable decision-making in the presence of quantum measurement noise.
Second, the precision-recall curves (Figure~\ref{fig:precision_recall_curve}) offer further insights into the model’s behavior under class imbalance. These curves highlight the trade-off between precision and recall, showing that the model remains effective in correctly identifying entangled states while maintaining robustness against false detections.
Finally, the scatter plot of predicted probabilities colored by prediction correctness (Figure~\ref{fig:scatter_probabilities}) visually demonstrates the model’s ability to assign higher confidence to correct predictions. This interpretability supports trust in the model's outputs and provides an intuitive understanding of its probabilistic behavior in realistic, noisy quantum settings.
These results, obtained under minimal training data and realistic noise assumptions, further support the practicality and robustness of the proposed method for entanglement detection.

Here, as an example, we analyze one of the EW constructed using ROEW, defined by the following matrix:

\begin{equation}  
	W = \scalebox{0.95}{$\begin{bmatrix}  
			0.459    & 0.019 - 0.002i & -0.002 - 0.019i & -0.002i \\
			0.019 + 0.002i & 0.055      & 0.399 - 0.012i & 0.017   \\
			-0.002 + 0.019i & 0.399 + 0.012i & 0.055     & -0.005 - 0.017i \\
			0.002i   & 0.017      & -0.005 + 0.017i & 0.430   
		\end{bmatrix}$},
	\label{eq:witness}  
\end{equation}

The witness matrix in~\eqref{eq:witness} (corresponding to the highlighted row in Table~\ref{tab:test}) is a Hermitian matrix with one negative eigenvalue. For the Bell state \( |\Psi^-\rangle = \frac{1}{\sqrt{2}} (|01\rangle - |10\rangle) \), we obtain:  
\begin{equation}  
	\text{tr}(W \rho_W) = -0.344 < 0,  
\end{equation}  
which confirms the detection of entanglement and for product states of the form:
\begin{equation}
	|\Phi_k\rangle = \cos\left(\tfrac{\theta_k}{2}\right) |0\rangle + e^{i \phi_k} \sin\left(\tfrac{\theta_k}{2}\right) |1\rangle, (k = A,B), \quad|\psi\rangle = |\Phi_A\rangle \otimes |\Phi_B\rangle, 
\end{equation}
then $ \text{tr}(W |\psi\rangle\langle \psi|)>0$ for all $\phi_k \in (0,2\pi)$ and $\theta_k \in (0,\pi)$, which confirms that $W$ is a valid EW.

\section{Conclusions}

 Our study highlights the potential of machine learning, particularly SVM, to classify quantum states effectively, even in the presence of noise. By introducing robust optimization protocols to develop a ROEW, we have demonstrated a method that enhances the resilience and efficiency of entanglement classification. The ROEW-based approach not only mitigates the impact of quantum measurement errors but also reduces the need for extensive feature sets, making it a resource-efficient solution. These advancements mark a significant step forward in overcoming noise challenges, paving the way for more reliable and practical quantum computing applications. A promising direction for future research involves extending the approach to multi-class classification tasks.

\begin{sidewaystable} 
	\centering
	\begin{tabular}{|c|cccccccccccccccc|}
		\hline
		$ \alpha $ & $\chi_{11}$ &$\chi_{12}$ &$\chi_{13}$ &$\chi_{14}$ &$\chi_{21}$ &$\chi_{22}$ &$\chi_{23}$ &$\chi_{24}$ &$\chi_{31}$ &$\chi_{32}$ &$\chi_{33}$ &$\chi_{34}$ &$\chi_{41}$ &$\chi_{42}$ &$\chi_{43}$ &$\chi_{44} (b^{opt}) $\\ 
		\hline 
		0.6   & -9.5  & -0.2  & 0.1  & 0.3  & -0.4  & 9.7  & -0.4  & 0.2  & -0.3  & -0.6  & -9.3  & 0.1  & -0.3  & 0.2  & -0.1 & 11.5 \\ 
		& 11.0  & 0.2  & 0.6  & 0.1  & -0.2  & -11.1  & -0.0  & -0.2  & 0.5  & -0.2  & -10.8  & 0.4  & 0.0  & 0.3  & -0.4 & 13.1  \\ 
		& -10.3  & -0.0  & -0.3  & -0.1  & 0.0  & -10.3  & -0.1  & -0.4  & -0.3  & -0.1  & 10.1  & -0.1  & 0.0  & 0.4  & -0.1 & 12.3 \\ 
		& 10.5  & -0.2  & -0.5  & 0.4  & 0.3  & 10.5  & 0.4  & 0.5  & 0.6  & -0.4  & 10.2  & 0.4  & 0.3  & 0.6  & 0.4 & 12.5 \\ 
		\hline
		0.7  & -7.6  & -0.1  & 0.1  & 0.2  & -0.3  & 7.8  & -0.3  & 0.1  & -0.2  & -0.5  & -7.4  & 0.0  & -0.2  & 0.1  & -0.0 & 9.7 \\ 
		& 8.6  & 0.1  & 0.4  & 0.1  & -0.1  & -8.6  & -0.0  & -0.1  & 0.4  & -0.2  & -8.4  & 0.3  & 0.0  & 0.2  & -0.3 & 10.7\\ 
		& -8.3  & 0.0  & -0.2  & -0.1  & -0.0  & -8.3  & -0.1  & -0.3  & -0.2  & -0.1  & 8.2  & -0.1  & -0.0  & 0.3  & -0.1  & 10.4 \\
		
		& \textbf{8.3}  & \textbf{-0.2}  & \textbf{-0.4}  & \textbf{0.3}  & \textbf{0.3}  & \textbf{8.3}  & \textbf{0.4}  & \textbf{0.4}  & \textbf{0.5}  & \textbf{-0.3}  & \textbf{8.1}  & \textbf{0.3}  & \textbf{0.3}  & \textbf{0.4}  & \textbf{0.3} & \textbf{10.4} \\ 
		\hline
		0.8   & -5.3  & -0.0  & 0.1  & 0.1  & -0.2  & 5.4  & -0.2  & 0.1  & -0.1  & -0.3  & -5.2  & 0.0  & -0.1  & 0.1  & 0.0  & 7.3\\ 
		& 5.9  & 0.1  & 0.3  & 0.1  & -0.1  & -5.9  & -0.0  & -0.1  & 0.3  & -0.1  & -5.8  & 0.2  & 0.0  & 0.2  & -0.2 & 7.9\\ 
		& -5.7  & 0.0  & -0.2  & -0.0  & -0.0  & -5.7  & -0.1  & -0.2  & -0.2  & -0.0  & 5.6  & -0.1  & -0.0  & 0.2  & -0.1 & 7.7 \\ 
		& 6.1  & -0.1  & -0.3  & 0.2  & 0.2  & 6.1  & 0.3  & 0.3  & 0.3  & -0.2  & 5.9  & 0.2  & 0.2  & 0.3  & 0.2 & 8.2 \\ 
		\hline
		0.9  & -3.7  & -0.0  & 0.0  & 0.0  & -0.1  & 3.7  & -0.1  & 0.0  & -0.1  & -0.2  & -3.6  & -0.0  & -0.1  & 0.0  & 0.0 & 5.7  \\ 
		& 4.1  & 0.1  & 0.2  & 0.0  & -0.1  & -4.1  & -0.0  & -0.1  & 0.2  & -0.1  & -4.0  & 0.1  & 0.0  & 0.1  & -0.1 & 6.2\\ 
		& -3.8  & 0.0  & -0.1  & 0.0  & -0.0  & -3.8  & -0.1  & -0.1  & -0.1  & -0.0  & 3.8  & -0.1  & -0.0  & 0.2  & -0.1 & 5.9 \\ 
		& 3.8  & -0.1  & -0.2  & 0.1  & 0.1  & 3.7  & 0.1  & 0.2  & 0.2  & -0.1  & 3.7  & 0.1  & 0.1  & 0.2  & 0.1 & 5.8\\ 
		\hline
	\end{tabular}
	\caption{Optimal values of Eq.\ref{eq:en} for different $\alpha$, where $\chi_{ij}$ sets the witness operator coefficients. }
	\label{tab:test}
\end{sidewaystable}

 \vspace{10cm}

\bibliographystyle{ieeetr}
\bibliography{reff}

\end{document}